\documentclass[aps, prx, reprint, superscriptaddress, showpacs, nofootinbib, longbibliography, floatfix]{revtex4-2}

\usepackage[utf8]{inputenc} 
\usepackage[english]{babel} 
\usepackage{booktabs} 
\usepackage{comment} 
\usepackage{hyperref} 
\usepackage{physics} 
\usepackage[table,xcdraw,dvipsnames]{xcolor} 
\allowdisplaybreaks 

\usepackage{lipsum} 
\usepackage{xfrac} 
\usepackage{multirow} 

\usepackage{amsmath}  
\usepackage{amssymb}  
\usepackage{mathrsfs} 
\usepackage{mathtools} 
\usepackage{amsthm}   
\usepackage{bbm}      
\usepackage{bm}       

\usepackage{indentfirst} 
\usepackage{latexsym}    

\usepackage{verbatim} 
\usepackage{cancel}   
\usepackage{url}      

\usepackage{subcaption} 

\usepackage[font=small, labelfont=bf, format=plain, justification=raggedright, singlelinecheck=false]{caption} 
\usepackage[title, titletoc]{appendix} 

\usepackage{graphicx} 
\graphicspath{{images/}} 
\usepackage{epstopdf} 

\newcommand{\caphead}[1]{{\bf #1}} 

\renewcommand{\thesection}{\Roman{section}}
\renewcommand{\thesubsection}{\Roman{section} \Alph{subsection}}
\renewcommand{\thesubsubsection}{\Roman{section} \Alph{subsection} \arabic{subsubsection}}

\makeatletter
\def\p@subsection{} 
\def\p@subsubsection{} 
\newcommand\footnoteref[1]{\protected@xdef\@thefnmark{\ref{#1}}\@footnotemark} 
\makeatother

\usepackage{soul} 

\usepackage{enumitem} 

\usepackage{qcircuit} 
\usepackage{hyperref}
\hypersetup{
  colorlinks=true,
  hypertexnames=false,
  linktocpage,
  colorlinks=true, 
  urlcolor=magenta!90!black,    
  linkcolor=blue!60!black, 
  citecolor=black!60 
}
\begin{document}

\title{Scalable and fault-tolerant preparation of encoded k-uniform states}

\author{Shayan Majidy}
\email{smajidy@fas.harvard.edu}
\affiliation{Department of Physics, Harvard University, Cambridge, Massachusetts 02138, USA}

\author{Dominik Hangleiter}
\affiliation{Simons Institute for the Theory of Computing, University of California, Berkeley, California 94720, USA}

\author{Michael J. Gullans}
\affiliation{Joint Center for Quantum Information and Computer Science, University of Maryland and NIST, College Park, Maryland 20742, USA}

\date{\today}

\begin{abstract} 
$k$-uniform states are valuable resources in quantum information, enabling tasks such as teleportation, error correction, and accelerated quantum simulations. The practical realization of $k$-uniform states, at scale, faces major obstacles: verifying $k$-uniformity is as difficult as measuring code distances, and devising fault-tolerant preparation protocols further adds to the complexity. To address these challenges, we present a scalable, fault-tolerant method for preparing encoded $k$-uniform states, and we illustrate our approach using surface and color codes. We first present a technique to determine $k$-uniformity of stabilizer states directly from their stabilizer tableau. We then identify a family of Clifford circuits that ensures both fault tolerance and scalability in preparing these states. Building on the encoded $k$-uniform states, we introduce a hybrid physical–logical strategy that retains some of the error-protection benefits of logical qubits while lowering the overhead for implementing arbitrary gates compared to fully logical algorithms. We show that this hybrid approach can outperform fully physical implementations for resource-state preparation, as demonstrated by explicit constructions of $k$-uniform states.
\end{abstract}

\maketitle

\section{Introduction}\label{sec:int}

A pure $N$-qubit state is called $k$-uniform if every subset of $k$ qubits is maximally mixed~\cite{scott2004multipartite, facchi2008maximally, arnaud2013exploring}. Such states are used in many quantum information processing tasks, including conventional and open-destination quantum teleportation~\cite{helwig2012absolute, helwig2013absolutely}, secret sharing~\cite{helwig2012absolute, helwig2013absolutely}, information masking~\cite{shi2021k}, and quantum error correction~\cite{raissi2020modifying, pastawski2015holographic, raissi2018optimal}. Well-known examples include the GHZ state ($k=1$), the five-qubit code ($k=2$)~\cite{laflamme1996perfect}, and the toric code ($k=3$)~\cite{dennis2002topological}.  More generally, any logical state of a stabilizer code with pure distance $d_p$ (minimum weight undetectable error \cite{wagner21}) is a $(d_p-1)$-uniform state. Beyond these direct applications, $k$-uniform states also appear in the black hole/qubit correspondence~\cite{borsten2010four, borsten2011black}, underlie states that can be employed as quantum repeaters~\cite{alsina2021absolutely}, and provide extreme realizations of local thermalization~\cite{popescu2006entanglement}, thereby further bridging thermodynamics and error correction~\cite{bilokur2024thermodynamic, landi2020thermodynamic}. They have also recently been shown to accelerate quantum simulations~\cite{zhao2024entanglement}.

Given their wide-ranging applications, considerable effort has been devoted to identifying $k$-uniform states. Techniques include brute-force numerical searches~\cite{borras2007multiqubit}, graph-state constructions~\cite{raissi2022general, sudevan2022n, raissi2020constructions, feng2017multipartite, helwig2013absolutely2, hein2004multiparty}, combinatorial design methods~\cite{shi2022k, zang2021quantum, li2019k, pang2019two, goyeneche2018entanglement, goyeneche2015absolutely, goyeneche2014genuinely}, and tools from statistical mechanics~\cite{facchi2010classical, di2018feynman}. In the same vein, there have been parallel efforts to construct proofs of both the existence and the non-existence of $k$-uniform states~\cite{ning2025linear, shi2024bounds, huber2017absolutely, huber2018bounds, burchardt2020stochastic}.

Despite these constructive advances, the practical realization of $k$-uniform states at scale remains challenging. The first major obstacle is verification as $N$ scales: confirming $k$-uniformity requires measuring the reduced density matrices of all $\binom{N}{k}$ subsets, a task equivalent to determining the distance of a quantum code—an NP-hard problem. The second challenge is fault-tolerant, logical state preparation: there is no general method for constructing fault-tolerant circuits that generate encoded $k$-uniform states. With quantum computing entering an era where logical qubits and fault-tolerant protocols are becoming experimentally viable~\cite{bluvstein2024logical, acharya2024quantum, google2023suppressing, krinner2022realizing, reichardt2024logical, rodriguez2024experimental, da2024demonstration, hong2024entangling, ryan2024high, majidy2024building}, addressing these challenges is now timely.

In this work, we present a numerical method to identify $k$-uniform stabilizer states which can be prepared fault-tolerantly in a natural way. Our approach leverages the stabilizer-tableau formalism to verify $k$-uniformity and systematically searches for families of Clifford circuits that are constant-depth, fault-tolerant, and scalable. We demonstrate our method in surface codes~\cite{fowler2012surface} and color codes~\cite{grassl1997codes}, providing practical schemes for preparing encoded $k$-uniform states with near-term quantum devices.

Building on the preparation of encoded $k$-uniform states, we propose a hybrid physical-logical approach to reduce the overhead of arbitrary-angle logical rotations while preserving error-protection benefits. Conventional methods such as code switching~\cite{anderson2014fault}, distillation~\cite{knill2004fault1, bravyi2005universal}, and cultivation~\cite{gidney2024magic} enable universal quantum computation but require substantial qubit and gate resources. Meanwhile, experimental platforms have achieved single-qubit gate fidelities exceeding 99.9\%, with some nearing 99.999\%~\cite{rower2024suppressing, loschnauer2024scalable, evered2023high}, suggesting that direct physical operations may be advantageous in certain regimes. To exploit this, we introduce a hybrid scheme that selectively unencodes qubits using mixed physical-logical Bell states, applies physical arbitrary-angle rotations, and then reencodes them. Through exact numerical simulations with realistic noise models, we identify scenarios where the benefits of unencoding outweigh its costs for state preparation. Our results show that this hybrid approach can yield higher-fidelity physical state preparation than purely physical circuits, as demonstrated in $k$-uniform state preparation.

The remainder of this manuscript is organized as follows. In Sec.~\ref{sec:numapproach}, we present our method for constructing scalable, fault-tolerant circuits for preparing encoded $k$-uniform states. Section~\ref{sec:kuniprotocol} demonstrates the application of this approach to surface and color codes. In Sec.~\ref{sec:hybridadvantage}, we explore how logical $k$-uniform states can enhance the fidelity of physical state preparation within our hybrid scheme. Finally, Sec.~\ref{sec:discussion} summarizes our findings and discusses future research directions.

\section{Circuit Construction Method}\label{sec:numapproach}

This section outlines our approach for identifying fault-tolerant, scalable circuits that prepare encoded $k$-uniform states. In Sec.~\ref{sub:calc}, we describe how to determine $k$-uniformity using a state's stabilizer tableau. Sec.~\ref{sub:general_circuit} then presents a circuit architecture designed for fault tolerance and scalability. A summary of our methodology is provided in Appendix~\ref{app:methodology}. We denote a quantum code with $n$ physical qubits, $\kappa$ logical qubits, and distance $d$ by $[n, \kappa, d]$, using $\kappa$ instead of $k$ to avoid confusion with $k$-uniformity. Additionally, we denote the total number of qubits in the state by $N$.

\subsection{Calculating \texorpdfstring{$k$}{k}-uniformity}\label{sub:calc}

Consider a stabilizer codespace for $k$ qubits, defined by $r$ stabilizer generators, which determines a set of $2^{k-r}$ codewords $\ket{s_i}$. The associated stabilizer mixed state is given by
\begin{equation} \rho = \frac{1}{2^{k-r}} \sum_i \dyad{s_i}.
\end{equation}
If $k = r$, $\rho$ is a pure stabilizer state. More generally, for any $N$-qubit stabilizer state $\ket{\psi}$ and any subset $A$ of qubits with $\abs{A} = k$, one can always find a set of $r$ stabilizer generators such that the reduced density matrix $\rho_A := \Tr_{\bar{A}}(\dyad{\psi})$ is itself a stabilizer mixed state~\cite{aaronson2004improved}.

To assess approximate $k$-uniformity, we employ the $\Delta$-approximate criterion from Ref.~\cite{zhao2024entanglement}. A state $\ket{\psi}$ is $\Delta$-approximate $k$-uniform if \begin{equation}
\| \Tr_{\bar{k}}[\dyad{\psi}] - I/2^k \|_1 \leq \Delta
\end{equation}
for every size-$k$ subset of qubits. In the stabilizer-mixed case, this simplifies to $|\rho_A - I/2^k|_1 = 2 - 2^{1-r} \leq \Delta$, so determining $\Delta$ for $\ket{\psi}$ and $k$ reduces to finding the maximum $r$ over all subsets $A$ with $\abs{A}=k$.

We now describe how to extract this $r$ from a stabilizer tableau. Let $I_A$ be the number of stabilizer generators that remain independent when their action on $\bar{A}$ is replaced by the identity. From Ref.~\cite{nahum2017quantum}, the von Neumann entropy of the reduced density matrix satisfies \begin{equation} S(\rho_A) = I_A - k. \end{equation} For stabilizer mixed states, $\rho_A$ is diagonal in an appropriate basis, giving $S(\rho_A) = k - r$. Equating these expressions yields
\begin{equation}
r = 2k - I_A.
\end{equation}
Thus, maximizing $r$ is equivalent to minimizing $I_A$ over all $\abs{A} = k$ subsets. This provides a direct way to quantify a state's $k$-uniformity from its stabilizer tableau. An example of this procedure is given in Appendix~\ref{app:kunicalc_example}.

\subsection{Scalable and Fault-Tolerant Architecture}\label{sub:general_circuit}

A na\"ive approach to generating $k$-uniform states is to randomly construct Clifford circuits and evaluate their $k$-uniformity using the method outlined above. However, such circuits typically lack fault tolerance and scalability, require long-range interactions, and exhibit large circuit depths. To address these issues, we restrict our search to a structured class of Clifford circuits.

\begin{figure}
\includegraphics[width=\columnwidth]{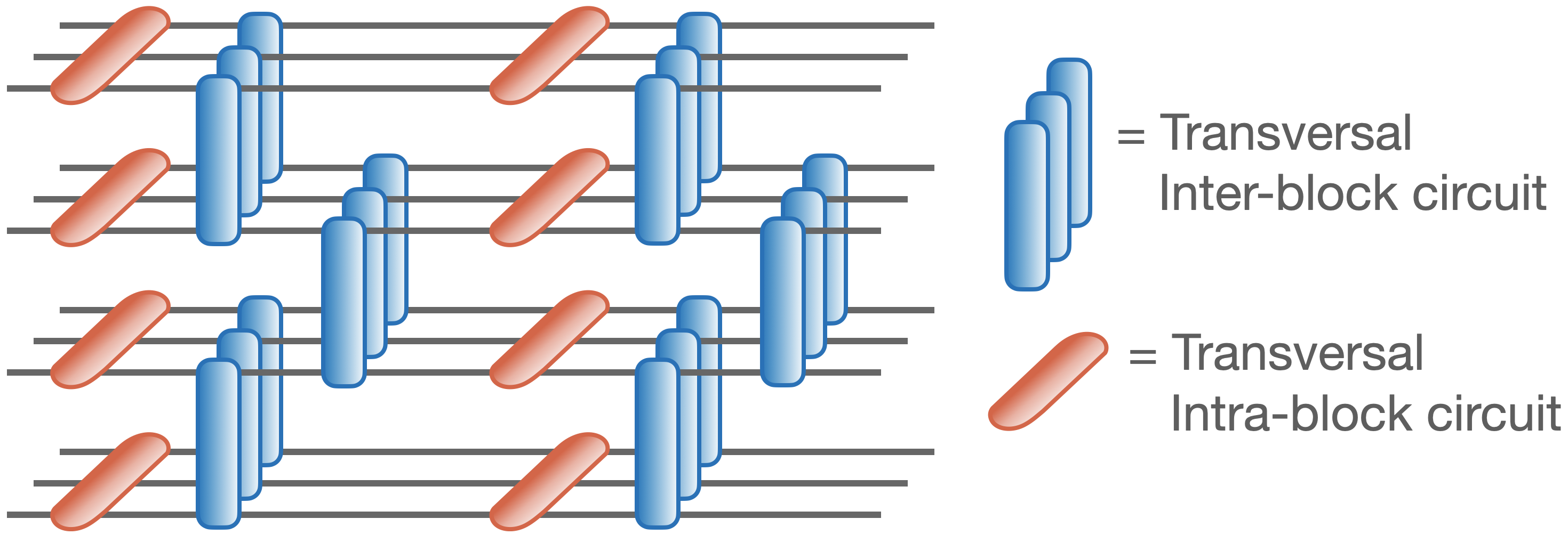}
\caption{\caphead{Fault-tolerant circuit architecture.} For an $[n, \kappa, d]$ code, logical qubits are grouped into $\kappa$-qubit blocks. Transversal gates are applied within each block, followed by two-qubit transversal gates between neighboring blocks in a brickwork pattern.}
 \label{fig:kuniprep}
\end{figure}

Figure~\ref{fig:kuniprep} illustrates our design for an $[n, \kappa=3, d]$ code. Logical qubits are grouped into blocks of size $\kappa$. The circuit first applies transversal gates within each block, followed by two-qubit transversal gates between adjacent blocks in a brickwork pattern. Because every gate is transversal, the circuit is inherently fault tolerant. Any product state that can be fault-tolerantly initialized serves as a suitable starting state, and we systematically search over circuits of this form.

To achieve scalability, we enforce spatially invariant circuit layers and refine boundary gates as needed. This guarantees a consistent bulk structure, allowing us to verify $k$-uniformity for large but finite $N$ and confidently extend it to even larger $N$. Additionally, we introduce the notion of $\alpha$-separated $k$-uniformity: a state is $\alpha$-separated $k$-uniform if all size-$k$ subsets of qubits spaced at least $\alpha$ sites apart satisfy $k$-uniformity. For example, for $k=4$ and $\alpha=3$, one checks subsets such as qubits 1, 4, 7, and 10. Once we identify a circuit that meets this reduced condition, we verify full $k$-uniformity by checking all remaining subsets, avoiding the need for an exhaustive $\Delta$ calculation at every step.

\section{Example Circuits}\label{sec:kuniprotocol}

We illustrate our method with two representative quantum codes. First, we examine the rotated surface code, a widely implemented architecture in recent quantum error correction experiments~\cite{acharya2024quantum, bluvstein2024logical, google2023suppressing, krinner2022realizing}. Second, we consider the $[4,2,2]$ 2D color code, which exemplifies cases where $\kappa>1$ and connects to multiple error-correcting code families, including hypercube quantum codes and hybrid classical-quantum codes~\cite{criger2016noise, kubica2015unfolding, majidy2023unification, renes2011efficient}. We validate our approach for systems of up to 40 qubits. Since the circuit bulk follows a uniform structure with only minor boundary modifications (e.g., reversing a small number of CNOT orientations), $k$-uniformity is preserved as $N$ increases.

\subsection{Surface Code} \label{sub:toriccodeeg}

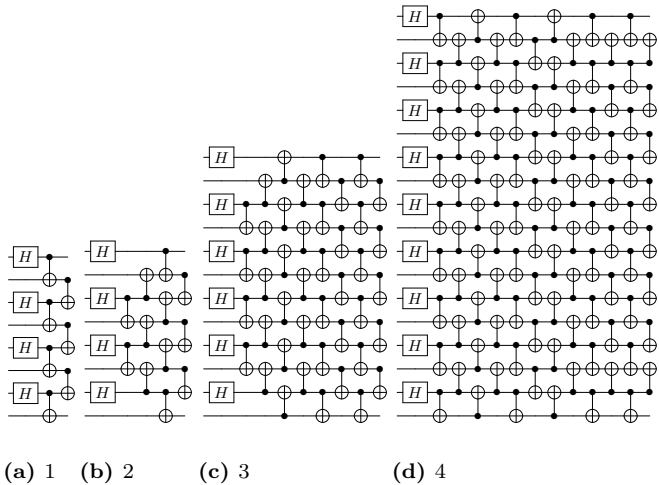
\begin{figure}
 \centering
 \begin{subfigure}[b]{0.105\columnwidth}
    \centering
    \begin{equation*}
    \scalebox{0.667}{
        \Qcircuit @C=0.3em @R=0.35em {
          & \gate{H} & \ctrl{1} & \qw \\
          & \qw & \targ & \ctrl{1}  \\
          & \gate{H} & \ctrl{1} & \targ  \\
          & \qw & \targ & \ctrl{1} \\
          & \gate{H} & \ctrl{1} & \targ \\
          & \qw & \targ & \ctrl{1} \\
          & \gate{H} & \ctrl{1} & \targ\\
          & \qw & \targ & \qw \\
        }
        }
    \end{equation*}
     \caption{\caphead{$1$}}
     \label{fig:k1_surface}
     \end{subfigure} 
     \begin{subfigure}[b]{0.17\columnwidth}
     \centering
    \begin{equation*}
     \scalebox{0.667}{
    \Qcircuit @C=0.35em @R=0.4em {
    & \gate{H} & \qw     & \qw       & \ctrl{1}  & \qw  \\
    & \qw      & \qw & \targ     & \targ     & \ctrl{1} \\
    & \gate{H} & \ctrl{1}  & \ctrl{-1} & \targ     & \targ \\
    & \qw      & \targ     & \targ     & \ctrl{-1} & \ctrl{1}  \\
    & \gate{H} & \ctrl{1}  & \ctrl{-1} & \targ     & \targ  \\
    & \qw      & \targ     & \targ     & \ctrl{-1} & \ctrl{1} \\
    & \gate{H} & \qw     & \ctrl{-1} & \ctrl{1}  & \targ  \\
    & \qw      & \qw & \qw       & \targ     & \qw     \\
    }
    }
    \end{equation*}
 \caption{\caphead{$2$}}
 \label{fig:k2_surface}
 \end{subfigure}
 \begin{subfigure}[b]{0.285\columnwidth}
 \centering
    \begin{equation*}
     \scalebox{0.667}{
    \Qcircuit @C=0.35em @R=0.4em {
    & \gate{H}  & \qw & \qw & \targ & \qw & \ctrl{1} & \qw & \ctrl{1} & \qw  \\
    & \qw       & \qw & \targ & \ctrl{-1} & \targ & \targ & \ctrl{1} & \targ & \ctrl{1} \\
    & \gate{H}  & \ctrl{1} & \ctrl{-1} & \targ & \ctrl{-1} & \ctrl{1} & \targ & \ctrl{1} & \targ \\
    & \qw       & \targ & \targ & \ctrl{-1} & \targ & \targ & \ctrl{1} & \targ & \ctrl{1} \\
    & \gate{H}  & \ctrl{1} & \ctrl{-1} & \targ & \ctrl{-1} & \ctrl{1} & \targ & \ctrl{1} & \targ \\
    & \qw       & \targ & \targ & \ctrl{-1} & \targ & \targ & \ctrl{1} & \targ & \ctrl{1} \\
    & \gate{H}  & \ctrl{1} & \ctrl{-1} & \targ & \ctrl{-1} & \ctrl{1} & \targ & \ctrl{1} & \targ \\
    & \qw       & \targ & \targ & \ctrl{-1} & \targ & \targ & \ctrl{1} & \targ & \ctrl{1} \\
    & \gate{H}  & \ctrl{1} & \ctrl{-1} & \targ & \ctrl{-1} & \ctrl{1} & \targ & \ctrl{1} & \targ \\
    & \qw       & \targ & \targ & \ctrl{-1} & \targ & \targ & \ctrl{1} & \targ & \ctrl{1} \\
    & \gate{H}  & \qw & \ctrl{-1} & \targ & \ctrl{-1} & \ctrl{1} & \targ & \ctrl{1} & \targ \\
    & \qw       & \qw & \qw & \ctrl{-1} & \qw & \targ & \qw & \targ & \qw \\
    }
    }
    \end{equation*}
 \caption{\caphead{$3$}}
 \label{fig:k3_surface}
 \end{subfigure}
 \begin{subfigure}[b]{0.4\columnwidth}
 \centering
    \begin{equation*}
     \scalebox{0.667}{
    \Qcircuit @C=0.35em @R=0.4em {
    & \gate{H}  & \ctrl{1} & \qw & \targ & \qw  & \ctrl{1} & \qw & \targ & \qw & \ctrl{1} & \qw & \ctrl{1} & \qw \\
    & \qw       & \targ & \targ & \ctrl{-1} & \targ & \targ & \ctrl{1} & \ctrl{-1} & \targ & \targ & \targ & \targ & \targ \\
    & \gate{H}  & \ctrl{1} & \ctrl{-1} & \targ & \ctrl{-1} & \ctrl{1} & \targ & \targ & \ctrl{-1} & \ctrl{1} & \ctrl{-1} &\ctrl{1} & \ctrl{-1} \\
    & \qw      & \targ & \targ & \ctrl{-1} & \targ & \targ & \ctrl{1} & \ctrl{-1} & \targ & \targ & \ctrl{1} &\targ & \ctrl{1} \\
    & \gate{H} & \ctrl{1} & \ctrl{-1} & \targ & \ctrl{-1} & \ctrl{1} & \targ & \targ & \ctrl{-1} & \ctrl{1} & \targ &\ctrl{1} & \targ \\
    & \qw      & \targ & \targ & \ctrl{-1} & \targ & \targ & \ctrl{1} & \ctrl{-1} & \targ & \targ & \ctrl{1} &\targ & \ctrl{1} \\
    & \gate{H} & \ctrl{1} & \ctrl{-1} & \targ & \ctrl{-1} & \ctrl{1} & \targ & \targ & \ctrl{-1} & \ctrl{1} & \targ &\ctrl{1} & \targ \\
    & \qw      & \targ & \targ & \ctrl{-1} & \targ & \targ & \ctrl{1} & \ctrl{-1} & \targ & \targ & \ctrl{1} &\targ & \ctrl{1} \\
    & \gate{H} & \ctrl{1} & \ctrl{-1} & \targ & \ctrl{-1} & \ctrl{1} & \targ & \targ & \ctrl{-1} & \ctrl{1} & \targ &\ctrl{1} & \targ \\
    & \qw      & \targ & \targ & \ctrl{-1} & \targ & \targ & \ctrl{1} & \ctrl{-1} & \targ & \targ & \ctrl{1} &\targ & \ctrl{1} \\
    & \gate{H} & \ctrl{1} & \ctrl{-1} & \targ & \ctrl{-1} & \ctrl{1} & \targ & \targ & \ctrl{-1} & \ctrl{1} & \targ &\ctrl{1} & \targ \\
    & \qw      & \targ & \targ & \ctrl{-1} & \targ & \targ & \ctrl{1} & \ctrl{-1} & \targ & \targ & \ctrl{1} &\targ & \ctrl{1} \\
    & \gate{H} & \ctrl{1} & \ctrl{-1} & \targ & \ctrl{-1} & \ctrl{1} & \targ & \targ & \ctrl{-1} & \ctrl{1} & \targ &\ctrl{1} & \targ \\
    & \qw      & \targ & \targ & \ctrl{-1} & \targ & \targ & \ctrl{1} & \ctrl{-1} & \targ & \targ & \ctrl{1} &\targ & \ctrl{1} \\
    & \gate{H} & \ctrl{1} & \ctrl{-1} & \targ & \ctrl{-1} & \ctrl{1} & \targ & \targ & \ctrl{-1} & \ctrl{1} & \targ &\ctrl{1} & \targ \\
    & \qw      & \targ & \targ & \ctrl{-1} & \targ & \targ & \ctrl{1} & \ctrl{-1} & \targ & \targ & \targ &\targ & \targ \\
    & \gate{H} & \ctrl{1} & \ctrl{-1} & \targ & \ctrl{-1} & \ctrl{1} & \targ & \targ & \ctrl{-1} & \ctrl{1} & \ctrl{-1} &\ctrl{1} & \ctrl{-1} \\
    & \qw     & \targ & \qw & \ctrl{-1} & \qw & \targ & \qw & \ctrl{-1} & \qw & \targ & \qw &\targ & \qw \\
    }
    }
    \end{equation*}
 \caption{\caphead{$4$}}
 \label{fig:k4_surface}
 \end{subfigure}
 \caption{\caphead{State preparation circuits for the surface code.} $k = $ 1 to 4 for plots (a) to (d). In all cases, the initial state is $\ket{0}^{\otimes N}$. Our protocol imposes the following requirements: (a) $N \geq 2$, (b) $N \geq 6$, (c) $N \geq 12$, and (d) $N \geq 18$.
 }
 \label{fig:rot_surface_eg}
\end{figure}

The rotated surface code has parameters $[L^2,1,L]$ and the logical $X$, $Z$, and CNOT gates are transversal. Here, $L$ is a free parameter that adjusts the code distance at the cost of additional physical qubits. A transversal Hadamard ($H$) can be enacted by physically rotating the code block~\cite{bluvstein2024logical}, but such mid-circuit rotations are often impractical, so we omit mid-circuit $H$ gates. Figure~\ref{fig:rot_surface_eg} shows four circuits our search produced for $k=1$ to 4, starting from the logical state $\ket{0}^{\otimes N}$. 

A CNOT with the control qubit on top is referred to here as ``upward,'' while one with the control qubit on the bottom is ``downward.'' For $k=1$, a single time step with two fully upward CNOT layers suffices. For $k=2$, the first and fourth CNOT layers are primarily upward, except for the outermost CNOTs of the first layer which are removed, while the second and third layers are primarily downward, except for the outermost CNOTs of the third layer which is downward. For $k=3$, the second to fourth layers are upward and the remaining layer are primarily upward, with the exception being the outermost CNOTs of the first layer which are removed. For $k=4$, layers 2, 3, 4, 7, and 8 are downward, with other layers upward except for boundary gates in layers 10 and 12. In each case, the central portion of the circuit has a consistent structure, while only a few boundary gates are flipped to ensure full $k$-uniformity for sufficiently large $N$. 

\subsection{Color Code}\label{sec:colorcodeg}

\begin{figure}
 \centering
 \begin{subfigure}[b]{0.14\columnwidth}
 \centering
 \begin{equation*}
  \scalebox{0.667}{
 \Qcircuit @C=0.35em @R=0.4em {
 & \ctrl{1} & \targ & \qw & \qw & \qw & \qw \\
 & \ctrl{-1} & \qw & \targ & \qw & \qw & \qw \\
 & \ctrl{1} & \ctrl{-2} & \qw & \targ & \qw & \qw\\
 & \ctrl{-1} & \qw & \ctrl{-2} & \qw & \targ & \qw \\
 & \ctrl{1} & \targ & \qw & \ctrl{-2} & \qw & \qw\\
 & \ctrl{-1} & \qw & \targ & \qw & \ctrl{-2} & \qw \\
 & \ctrl{1} & \ctrl{-2} & \qw & \targ & \qw & \qw\\
 & \ctrl{-1} & \qw & \ctrl{-2} & \qw & \targ & \qw \\
 & \ctrl{1} & \qw & \qw & \ctrl{-2} & \qw & \qw \\
 & \ctrl{-1} & \qw & \qw & \qw & \ctrl{-2} & \qw\\
 } 
 }
 \end{equation*}
 \caption{\caphead{$1$}}
 \label{fig:k1}
 \end{subfigure} 
 \begin{subfigure}[b]{0.14\columnwidth}
 \centering
 \begin{equation*}
  \scalebox{0.667}{
 \Qcircuit @C=0.35em @R=0.4em {
 & \ctrl{1} & \targ & \qw & \qw & \qw & \qw \\
 & \ctrl{-1} & \qw & \targ & \qw & \qw & \qw \\
 & \ctrl{1} & \ctrl{-2} & \qw & \ctrl{2} & \qw & \qw \\
 & \ctrl{-1} & \qw & \ctrl{-2} & \qw & \ctrl{2} & \qw \\
 & \ctrl{1} & \targ & \qw & \targ & \qw & \qw \\
 & \ctrl{-1} & \qw & \targ & \qw & \targ & \qw \\
 & \ctrl{1} & \ctrl{-2} & \qw & \ctrl{2} & \qw & \qw \\
 & \ctrl{-1} & \qw & \ctrl{-2} & \qw & \ctrl{2} & \qw \\
 & \ctrl{1} & \targ & \qw & \targ & \qw & \qw \\
 & \ctrl{-1} & \qw & \targ & \qw & \targ & \qw \\
 & \ctrl{1} & \ctrl{-2} & \qw & \qw & \qw & \qw \\
 & \ctrl{-1} & \qw & \ctrl{-2} & \qw & \qw & \qw \\
 } 
 }
 \end{equation*}
 \caption{\caphead{$2$}}
 \label{fig:k2}
 \end{subfigure}
 \begin{subfigure}[b]{0.28\columnwidth}
 \centering
 \begin{equation*}
  \scalebox{0.667}{
 \Qcircuit @C=0.35em @R=0.4em {
 & \ctrl{1} & \ctrl{2} & \qw & \qw & \qw & \ctrl{1} & \ctrl{2} & \qw & \qw & \qw \\
 & \ctrl{-1} & \qw & \ctrl{2}& \qw & \qw & \ctrl{-1} & \qw & \ctrl{2} & \qw & \qw & \qw \\
 & \ctrl{1} & \targ & \qw & \ctrl{2} & \qw & \ctrl{1} & \targ & \qw & \ctrl{2} & \qw & \qw \\
 & \ctrl{-1} & \qw & \targ & \qw & \ctrl{2} & \ctrl{-1} & \qw & \targ & \qw & \ctrl{2} & \qw \\
 & \ctrl{1} & \targ & \qw & \targ & \qw & \ctrl{1} & \targ & \qw & \targ & \qw & \qw \\
 & \ctrl{-1} & \qw & \targ & \qw & \targ & \ctrl{-1} & \qw & \targ & \qw & \targ & \qw \\
 & \ctrl{1} & \ctrl{-2} & \qw & \targ & \qw & \ctrl{1} & \ctrl{-2} & \qw & \targ & \qw & \qw \\
 & \ctrl{-1} & \qw & \ctrl{-2} & \qw & \targ & \ctrl{-1} & \qw & \ctrl{-2} & \qw & \targ & \qw \\
 & \ctrl{1} & \targ & \qw & \ctrl{-2} & \qw & \ctrl{1} & \targ & \qw & \ctrl{-2} & \qw & \qw\\
 & \ctrl{-1} & \qw & \targ & \qw & \ctrl{-2} & \ctrl{-1} & \qw & \targ & \qw & \ctrl{-2} & \qw \\
 & \ctrl{1} & \ctrl{-2} & \qw & \qw & \qw & \ctrl{1} & \ctrl{-2} & \qw & \qw & \qw & \qw \\
 & \ctrl{-1} & \qw & \ctrl{-2} & \qw & \qw & \ctrl{-1} & \qw & \ctrl{-2} & \qw & \qw & \qw \\
 } 
 }
 \end{equation*}
 \caption{\caphead{$3$}}
 \label{fig:k3}
 \end{subfigure}
 \begin{subfigure}[b]{0.4\columnwidth}
 \centering
\begin{equation*}
 \scalebox{0.667}{
\Qcircuit @C=0.35em @R=0.4em {
& \ctrl{1} & \ctrl{2} & \qw & \qw & \qw & \ctrl{1} & \ctrl{2} & \qw & \qw & \qw & \ctrl{1} & \ctrl{2} & \qw & \qw & \qw & \qw \\
& \ctrl{-1} & \qw & \ctrl{2}& \qw & \qw & \ctrl{-1} & \qw & \ctrl{2}& \qw & \qw & \ctrl{-1} & \qw & \ctrl{2}& \qw & \qw & \qw \\
& \ctrl{1} & \targ & \qw & \ctrl{2} & \qw & \ctrl{1} & \targ & \qw & \ctrl{2} & \qw & \ctrl{1} & \targ & \qw & \ctrl{2} & \qw & \qw \\
& \ctrl{-1} & \qw & \targ & \qw & \ctrl{2} & \ctrl{-1} & \qw & \targ & \qw & \ctrl{2} & \ctrl{-1} & \qw & \targ & \qw & \ctrl{2} & \qw \\
& \qw & \ctrl{2} & \qw & \targ & \qw & \qw & \ctrl{2} & \qw & \targ & \qw & \qw & \ctrl{2} & \qw & \targ & \qw & \qw \\
& \qw & \qw & \ctrl{2} & \qw & \targ & \qw & \qw & \ctrl{2} & \qw & \targ & \qw & \qw & \ctrl{2} & \qw & \targ & \qw \\
& \ctrl{1} & \targ & \qw & \ctrl{2} & \qw & \ctrl{1} & \targ & \qw & \ctrl{2} & \qw & \ctrl{1} & \targ & \qw & \ctrl{2} & \qw & \qw \\
& \ctrl{-1} & \qw & \targ & \qw & \ctrl{2} & \ctrl{-1} & \qw & \targ & \qw & \ctrl{2} & \ctrl{-1} & \qw & \targ & \qw & \ctrl{2} & \qw \\
& \ctrl{1} & \ctrl{2} & \qw & \targ & \qw & \ctrl{1} & \ctrl{2} & \qw & \targ & \qw & \ctrl{1} & \ctrl{2} & \qw & \targ & \qw & \qw \\
& \ctrl{-1} & \qw & \ctrl{2} & \qw & \targ & \ctrl{-1} & \qw & \ctrl{2} & \qw & \targ & \ctrl{-1} & \qw & \ctrl{2} & \qw & \targ & \qw \\
& \ctrl{1} & \targ & \qw & \targ & \qw & \ctrl{1} & \targ & \qw & \targ & \qw & \ctrl{1} & \targ & \qw & \targ & \qw & \qw \\
& \ctrl{-1} & \qw & \targ & \qw & \targ & \ctrl{-1} & \qw & \targ & \qw & \targ & \ctrl{-1} & \qw & \targ & \qw & \targ& \qw \\
& \ctrl{1} & \targ & \qw & \ctrl{-2} & \qw & \ctrl{1} & \targ & \qw & \ctrl{-2} & \qw & \ctrl{1} & \targ & \qw & \ctrl{-2} & \qw & \qw \\
& \ctrl{-1} & \qw & \targ & \qw & \ctrl{-2} & \ctrl{-1} & \qw & \targ & \qw & \ctrl{-2} & \ctrl{-1} & \qw & \targ & \qw & \ctrl{-2} & \qw \\
& \ctrl{1} & \ctrl{-2} & \qw & \targ & \qw & \ctrl{1} & \ctrl{-2} & \qw & \targ & \qw & \ctrl{1} & \ctrl{-2} & \qw & \targ & \qw & \qw \\
& \ctrl{-1} & \qw & \ctrl{-2} & \qw & \targ & \ctrl{-1} & \qw & \ctrl{-2} & \qw & \targ & \ctrl{-1} & \qw & \ctrl{-2} & \qw & \targ & \qw \\
& \ctrl{1} & \targ & \qw & \ctrl{-2} & \qw & \ctrl{1} & \targ & \qw & \ctrl{-2} & \qw & \ctrl{1} & \targ & \qw & \ctrl{-2} & \qw & \qw \\
& \ctrl{-1} & \qw & \targ & \qw & \ctrl{-2} & \ctrl{-1} & \qw & \targ & \qw & \ctrl{-2} & \ctrl{-1} & \qw & \targ & \qw & \ctrl{-2} & \qw \\
& \qw & \ctrl{-2} & \qw & \qw & \qw & \qw & \ctrl{-2} & \qw & \qw & \qw & \qw & \ctrl{-2} & \qw & \qw & \qw & \qw \\
& \qw & \qw & \ctrl{-2} & \qw & \qw & \qw & \qw & \ctrl{-2} & \qw & \qw & \qw & \qw & \ctrl{-2} & \qw & \qw & \qw \\
} 
}
\end{equation*}
 \caption{\caphead{$4$}}
 \label{fig:k4}
 \end{subfigure}
 \caption{\caphead{State preparation circuits for the $[4,2,2]$ color code.} $k = $ 1 to 4 for plots (a) to (d). In all cases, the initial state is $\ket{+}^{\otimes N}$. Our protocol imposes the following requirements:  (a) $N \geq 2$, (b) $N \geq 6$, (c) $N \geq 10$, and (d) $N \geq 20$. For case (d), if $N/2$ is odd, the second-to-last CZ gate is omitted instead of the last one.
 }
 \label{fig:color_eg}
\end{figure}
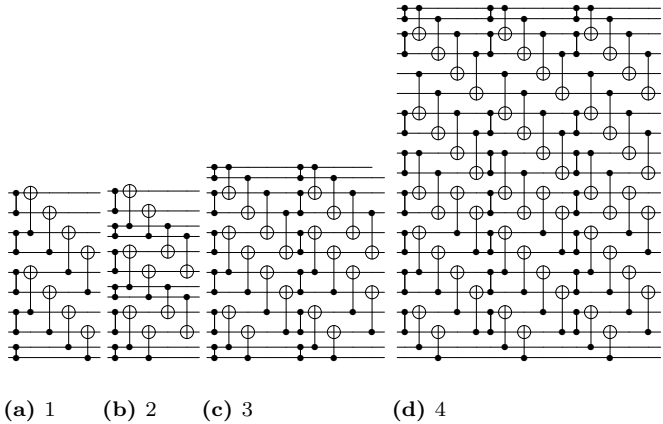

To illustrate a code with $\kappa>1$, we turn to the [4,2,2] 2D color code. This code supports transversal in-block $X$, $Z$, and CZ gates, while inter-block operations are realized via CNOTs. We focus on circuits that begin in the product state $\ket{+}^{\otimes N}$. Higher connectivity allows for shorter circuit depths than in the surface code example. Exact $k$-uniform circuits for $k\le 4$ appear in Fig.~\ref{fig:color_eg}; they again share a uniform bulk structure with slight boundary gate modifications. For $k=1$, all CZ gates appear within each block, and all CNOTs are oriented upward. For $k=2$, we start with the $k=1$ circuit and add a layer of downward-directed CNOTs. For $k=3$, the circuit is constructed from two copies of the $k=1$ architecture, with the first four rows of CNOTs reversed. For $k=4$, it is formed by combining three copies of the $k=1$ circuit, reversing the first ten rows of CNOTs, omitting the third CZ gate, and removing either the last or second-to-last CZ gate, depending on the parity of $N/2$.

Beyond exact $k$-uniform states, $\Delta$-approximate $k$-uniform states can be advantageous in applications such as product-formula simulations~\cite{zhao2024entanglement}. We set $\Delta=1$ (the smallest non-zero value) to focus on larger $k$. In particular: For $k=5$, repeating one time step of the $k=3$ circuit three times suffices for $N \ge 20$. For $k=6$, five repetitions of one time step from the $k=1$ circuit work for $N \ge 24$. For $k=7$, five repetitions of the $k=3$ circuit suffice for $n \ge 32$. These smaller circuit depths reflect the relaxed requirement of $\Delta$-approximate uniformity. Notably, increasing the depth of these circuits further does not continue to improve $k$-uniformity (see Appendix~\ref{app:deltaapprox}). 

\section{Hybrid Scheme and Physical State Preparation}\label{sec:hybridadvantage}

We now build on the circuits introduced in the previous section to explore the hybrid scheme. In Sec.~\ref{sub:hyb}, we formally present this scheme, followed by a numerical comparison with a purely physical implementation in Sec.~\ref{sub:numcomp}. We do not compare against a fully logical, fault-tolerant scheme, as the hybrid approach is only relevant in resource-limited scenarios where such a scheme would be impractical. To maintain near-term feasibility, we focus on small-distance codes.

\subsection{Hybrid Scheme}\label{sub:hyb}

We illustrate this scheme using a single logical qubit encoded in an $[n, \kappa = 1, d]$ code with logical operators $\overline{X}$ and $\overline{Z}$. The process begins by preparing a mixed logical–physical Bell state:
\begin{equation}
    \frac{1}{\sqrt{2}} \bigl(\ket{\overline{0}0} + \ket{\overline{1}1}\bigr),
\end{equation}
where the bar indicates that the qubit is logical.

There are multiple ways to generate this mixed Bell state. One method constructs a circuit that prepares $\ket{\overline{+}}$ and couples it to a physical qubit via $d$ CNOT gates, where the CNOT control qubits correspond to a representation of an $X$-type logical operator, and each target is the physical qubit. Circuit synthesis techniques can then minimize the number of operations on the physical qubit. Alternatively, one can prepare $\ket{\overline{+}+}$ and perform a $\overline{Z}Z$ measurement, postselecting on the desired outcome to obtain the target state. Another method first prepares a physical Bell state and then encodes one of its qubits, using, for instance, the technique described in Ref.~\cite{li2015magic}.

\begin{figure}
    \centering
    \includegraphics[width=\columnwidth]{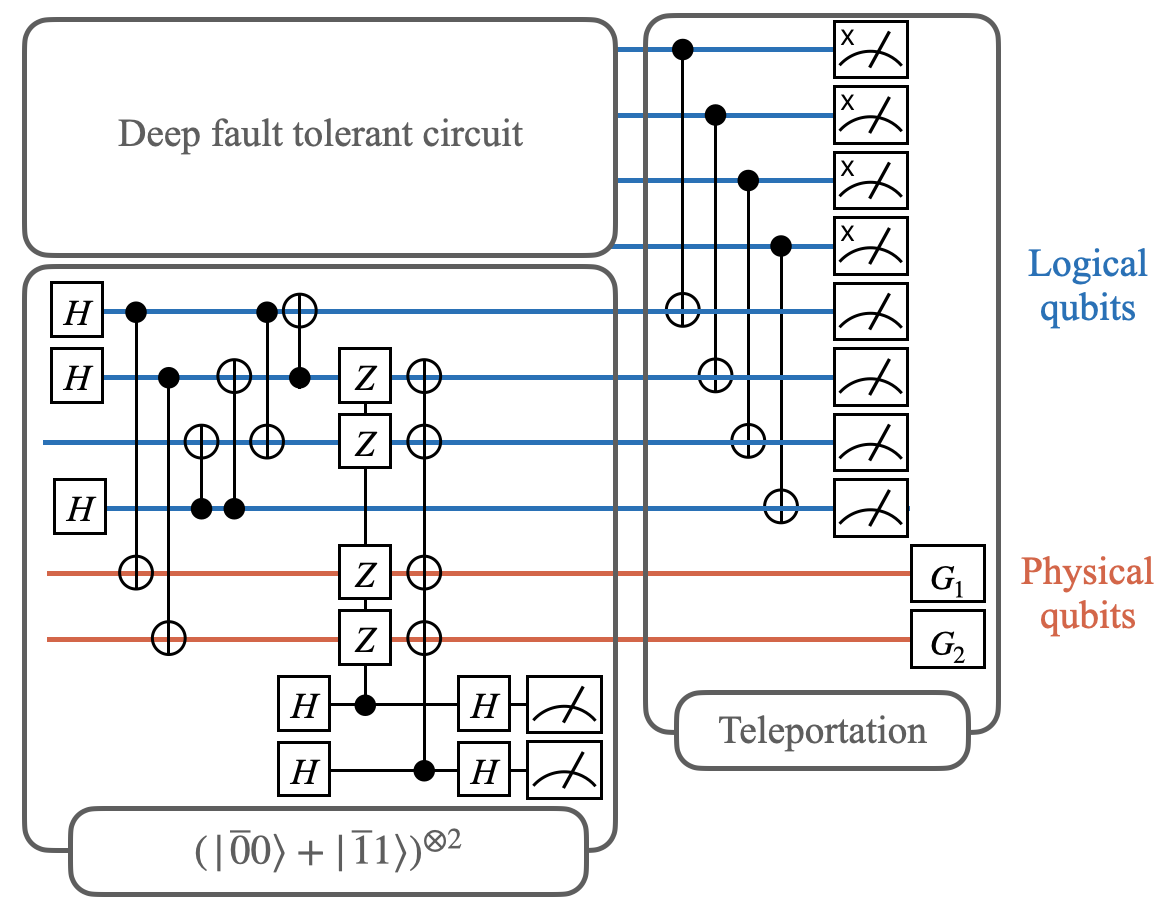}
    \caption{\caphead{Example of the unencoding protocol for the $[4,2,2]$ color code.} The protocol begins with the preparation of the teleportation resource state, $\tfrac{1}{\sqrt{2}}\left(\ket{\overline{0}0} + \ket{\overline{1}1}\right)$. Logical qubits are manipulated using their logical gate counterparts during the teleportation process. Correction gates $G_1$ and $G_2$ are applied based on measurement outcomes to complete the protocol.}
    \label{fig:teleport_eg}
\end{figure}

After preparing the mixed Bell state, we perform standard quantum teleportation. Specifically, we apply a CNOT between logical code blocks, measure all logical qubits in either the $X$- or $Z$-basis, and apply feed-forward corrections on the physical qubit as needed. Figure~\ref{fig:teleport_eg} illustrates this process. Notably, this procedure can switch between any two codes, not just between a code and no code.

We present a few examples of circuits that prepare mixed Bell states using the first method described. In each example, the final qubits represent the physical ones. For the $[7,1,3]$ code~\cite{steane1996multiple}, we simplified the circuit to require only a single gate on the physical qubit.
\begin{equation}
 \scalebox{0.66}{
        \Qcircuit @C=0.75em @R=0.75em  {
        & \gate{H} & \ctrl{7} & \qw & \qw & \ctrl{6} & \qw & \targ & \qw & \qw \\
        & \gate{H} & \qw & \ctrl{4} &\qw & \qw & \ctrl{3} &\ctrl{-1} & \targ & \qw\\
        & \gate{H} & \qw & \qw & \ctrl{4} & \qw & \qw & \qw & \ctrl{-1} & \qw\\
        & \gate{H} & \qw & \qw & \qw & \qw & \qw & \ctrl{3} & \ctrl{1} & \qw\\
        & \qw & \qw & \qw & \qw & \qw & \targ &\qw & \targ & \qw\\
        & \qw & \qw & \targ & \qw &\qw &\qw &\qw &\targ & \qw\\
        & \qw & \qw & \qw  &\targ & \targ &\qw &\targ &\ctrl{-1} & \qw\\
        & \qw & \targ & \qw &\qw &\qw &\qw &\qw &\qw  & \qw \\
        }}
\end{equation}
For the $[9,1,3]$ surface code, we also achieved a circuit with one CNOT on the physical qubit:
\begin{equation}
        \scalebox{0.66}{
        \Qcircuit @C=0.75em @R=0.75em  {
        & \gate{H} & \ctrl{1} & \qw  & \qw & \qw & \qw & \qw & \qw \\
        & \qw & \targ  & \qw  & \qw & \qw & \targ & \qw & \qw \\
        & \gate{H} & \ctrl{1} & \qw & \ctrl{6}& \qw & \qw & \qw & \qw  \\
        & \qw & \targ & \qw & \qw & \qw & \qw & \qw & \qw \\
        & \gate{H} & \ctrl{1} & \qw & \qw & \ctrl{3} & \qw & \ctrl{4} & \qw \\
        & \qw & \targ & \qw & \qw  & \qw & \qw &\qw & \qw  \\
        & \gate{H} & \ctrl{1}& \qw & \qw & \qw & \qw &\qw & \qw \\
        & \qw & \targ & \targ  & \qw & \targ & \qw &\qw & \qw \\
        & \gate{H} & \ctrl{1} &\ctrl{-1} &\targ & \qw & \ctrl{-7} & \targ & \qw \\
        & \qw & \targ & \qw & \qw  & \qw & \qw &\qw & \qw  \\
        }}
\end{equation}
For $\kappa > 1$, we need to create multiple pairs of Bell states. For instance, with the $[4,2,2]$ code, we prepare
\begin{equation}
\frac{1}{4}\left(\ket{\overline{00}00} + \ket{\overline{01}01} + \ket{\overline{10}10} + \ket{\overline{11}11}\right) 
\end{equation}
by first creating $\ket{\overline{++}}$ and then coupling each physical qubit with $d$ CNOTs, followed by circuit simplification to achieve one CNOT per physical qubit:
\begin{equation}
        \scalebox{0.66}{
        \Qcircuit @C=0.75em @R=0.75em  {
          & \gate{H} & \ctrl{4} & \qw & \qw & \qw & \ctrl{2} & \targ &  \qw & \qw & \qw \\
          & \gate{H} & \qw & \ctrl{4} & \qw  & \targ & \qw & \ctrl{-1}& \gate{Z} & \gate{X} & \qw \\
          & \qw      & \qw & \qw  & \targ & \qw & \targ &\qw & \gate{Z} \qwx[-1] & \gate{X} \qwx[-1] & \qw\\
          & \gate{H} & \qw & \qw & \ctrl{-1} & \ctrl{-2} &\qw & \qw &\qw &\qw & \qw\\
          & \qw      & \targ & \qw & \qw & \qw & \qw  & \qw & \gate{Z} \qwx[-2] & \gate{X} \qwx[-2]  & \qw \\
          & \qw      & \qw & \targ & \qw  & \qw & \qw  & \qw & \gate{Z} \qwx[-1] & \gate{X} \qwx[-1]  & \qw \\
          & & & & & & & \gate{H} & \ctrl{-1} & \qw & \gate{H} &\meter \\
          & & & & & & & \gate{H} & \qw & \ctrl{-2} & \gate{H} & \meter
        }}
\end{equation}
New stabilizers can be introduced to track how errors propagate, which is presented in the above circuit. For example, an $X$ error on the first qubit at the end of the circuit is detected after teleportation because it moves the logical state out of the code space, whereas an $X$ error at the beginning may remain unnoticed by the teleportation measurements but still be flagged by an additional stabilizer.

\subsection{Numerical Comparison}\label{sub:numcomp}

\begin{figure}[t]
\begin{subfigure}[b]{0.4925\columnwidth}
    \centering
    \includegraphics[width=\columnwidth]{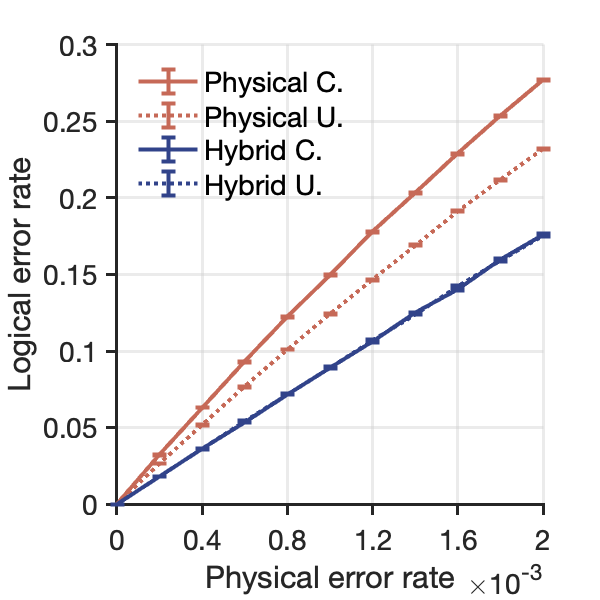}
    \caption{\caphead{GHZ preparation}}
    \label{fig:QEDa}
    \end{subfigure}
    \begin{subfigure}[b]{0.4925\columnwidth}
    \centering
    \includegraphics[width=\columnwidth]{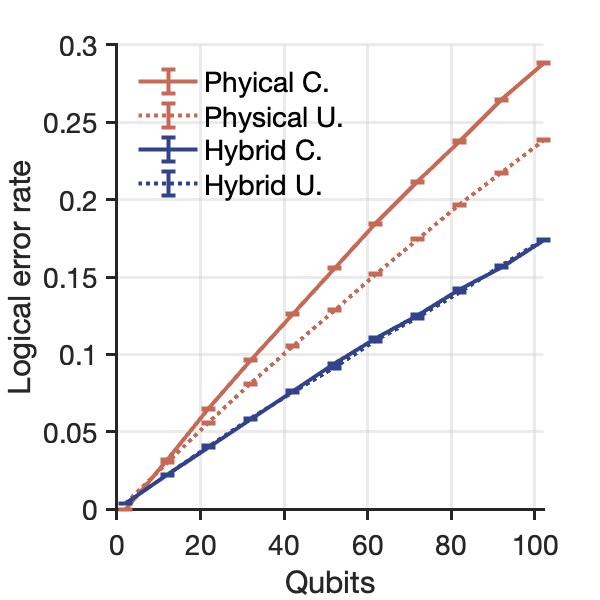}
    \caption{\caphead{GHZ preparation}}
    \label{fig:QEDb}
    \end{subfigure}
    \begin{subfigure}[b]{0.4925\columnwidth}
    \centering
    \includegraphics[width=\columnwidth]{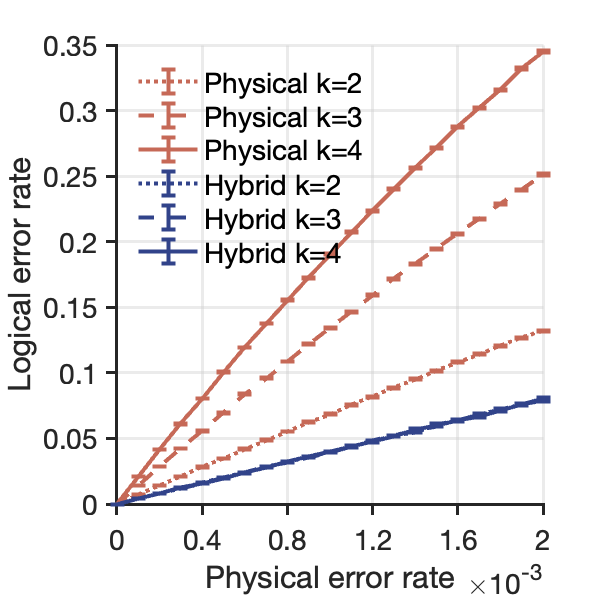}
    \caption{\caphead{$k$-uni preparation}}
    \label{fig:QEDc}
    \end{subfigure}
    \begin{subfigure}[b]{0.4925\columnwidth}
    \centering
    \includegraphics[width=\columnwidth]{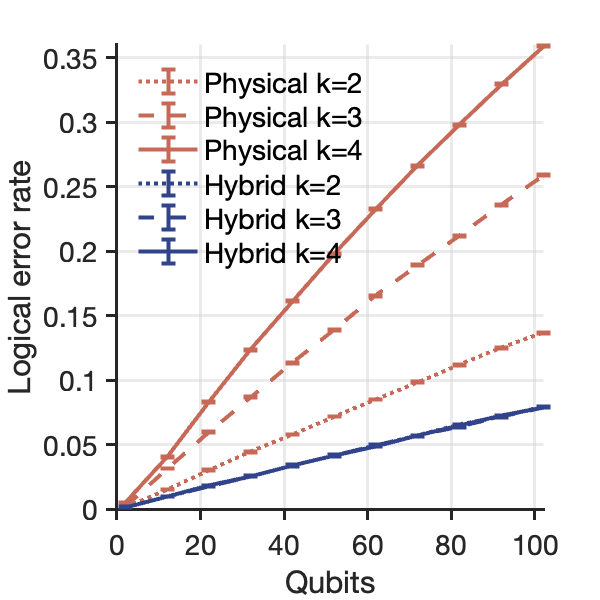}
    \caption{\caphead{$k$-uni preparation}}
    \label{fig:QEDd}
    \end{subfigure}
    \caption{\caphead{Hybrid circuits outperforming physical circuits in state preparation even when using small distance codes.} Numerical comparison of the hybrid and physical circuits in preparing physical $k$-uniform states under realistic noise models. For all plots we use $10^{6}$ samples of the circuit. The error bars are included in both figures but are too small to be clearly visible. For (a) to (c) we set qubits = $50$, $p_0 = p/100$, $p_1 = p/10$, and $p_2 = p_3 = p$. $p$ is plotted on the $x$-axis. For (d) to (f) We set $p_0 = 10^{-5}$, $p_1 = 10^{-4}$, $p_2 = 10^{-3}$, and $p_3 = 10^{-3}$.
    }
    \label{fig:hybrid_phs_compare}
\end{figure} 
The relative performance of the physical and hybrid schemes will naturally depend on the target state, its preparation method, and the choice of error-correcting code. To understand their relative advantages, we consider a few scenarios, beginning with one that favours the physical approach. We then examine more general cases, exploring quantum error detection and correction. Through out this section, we model noise using one- and two-qubit depolarizing channels with error rates: $p_0$ (idle), $p_1$ (single-qubit gates), $p_2$ (two-qubit gates), and $p_3$ (measurements). 

To favour the physical approach we prepare a $k=1$-uniform state using a minimum distance ($d=2$) code. Furthermore, we study a $k=1$ state whose circuits have been extensively optimized, the GHZ state. We consider two methods for preparing GHZ states. The first uses a constant-depth circuit with measurements and feed-forward. For example, the 6-qubit GHZ state preparation circuit is:
\begin{equation}
\scalebox{0.66}{
 \Qcircuit @C=0.2em @R=0.2em {
 & \gate{H} & \ctrl{1} & \ctrl{2} & \qw & \qw & \qw & \qw & \gate{X} \cwx[1] & \qw \\
 & \qw & \targ & \qw & \ctrl{2} & \qw & \qw & \qw & \gate{X} \cwx[1] & \qw \\
 & \qw & \qw & \targ & \qw & \targ & \qw & \meter &\cw \cwx[1]\\ 
 & \qw & \qw & \qw & \targ & \qw & \targ & \meter &\cw \cwx[1]\\
 & \gate{H} & \ctrl{1} & \ctrl{2} & \qw & \ctrl{-2} & \qw & \qw & \gate{X} \cwx[1] & \qw \\
 & \qw & \targ & \qw & \ctrl{2} & \qw & \ctrl{-2} & \qw & \gate{X} \cwx[1] & \qw \\ 
 & \qw & \qw & \targ & \qw & \targ & \qw & \meter &\cw \cwx[1] \\
 & \qw & \qw & \qw & \targ & \qw & \targ & \meter &\cw \\
 & \gate{H} & \ctrl{1} & \qw & \qw & \ctrl{-2} & \qw & \qw & \qw & \qw \\
 & \qw & \targ & \qw & \qw & \qw & \ctrl{-2} & \qw & \qw & \qw 
 }
 }
\end{equation}
The second method avoids measurements and classical feed-forward corrections, but requires a deeper (logarithmic-depth) circuit. Below is an example circuit for an 8-qubit GHZ state:
\begin{equation}
\scalebox{0.66}{
 \Qcircuit @C=0.2em @R=0.2em {
 & \gate{H} & \ctrl{1} & \ctrl{2} & \qw & \ctrl{4} & \qw & \qw & \qw & \qw\\
 & \qw & \targ & \qw & \ctrl{2} & \qw & \ctrl{4} & \qw & \qw & \qw\\
 & \qw & \qw & \targ & \qw & \qw & \qw & \ctrl{4} & \qw & \qw \\
 & \qw & \qw & \qw & \targ & \qw & \qw & \qw & \ctrl{4} & \qw\\
 & \qw & \qw & \qw & \qw & \targ & \qw & \qw& \qw & \qw\\
 & \qw & \qw & \qw & \qw & \qw & \targ & \qw& \qw & \qw\\
 & \qw & \qw & \qw & \qw & \qw & \qw & \targ& \qw & \qw\\
 & \qw & \qw & \qw & \qw & \qw & \qw & \qw& \targ & \qw
 } 
 }
\end{equation}

Figures~\ref{fig:QEDa} and~\ref{fig:QEDb} show the logical fidelity of these circuits as a function of qubit count and physical error rates for the hybrid and physical circuits. For the physical circuits, the constant-depth method results in lower fidelity than the log-depth due to its higher number of physical gates per qubit. The hybrid circuits perform similarly, as both approaches are constrained by the same dominant error source—the non-fault-tolerant teleportation protocol. The hybrid scheme outperforms the physical one in both cases, with the advantage increasing as the error rate rises for a fixed qubit number and as the qubit number grows for a fixed error rate.

Next, we extend our comparison to states with higher $k$, which require deeper circuits for physical preparation, as shown in Figure~\ref{fig:color_eg}. Figures~\ref{fig:QEDc} and~\ref{fig:QEDd} compare the resulting logical fidelity. In these cases, the hybrid method outperforms purely physical preparation. The key distinction lies in how error rates scale with increasing $k$. Physical circuits accumulate errors as depth increases, leading to progressively worse fidelity. Hybrid circuits maintain nearly constant error rates, since the limiting factor is the single non-fault-tolerant teleportation unencoding step. Once this step is performed, additional operations for increasing $k$-uniformity do not introduce proportionally higher errors. This result highlights a fundamental advantage of hybrid approaches: they scale more robustly with circuit depth and complexity than purely physical implementations. 

\begin{figure}[t]
    \centering
    \includegraphics[width=0.75\columnwidth]{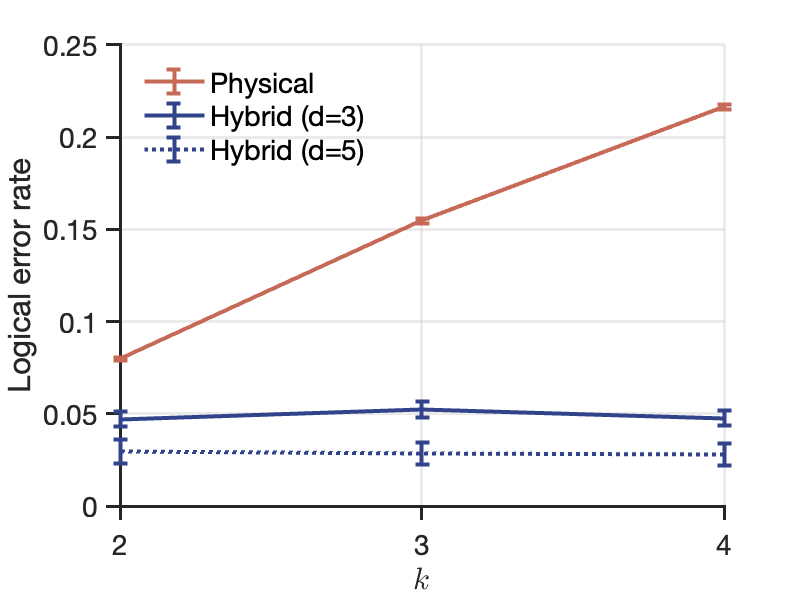}
    \caption{\caphead{Physical error rate increases with $k$, while Hybrid error rate remains constant using $[L^2,1,L]$ surface code.} Numerical comparison of the hybrid and physical circuit in preparing physical $k$-uniform states under realistic noise models. We set $p_0 = 10^{-5}$, $p_1 = 10^{-4}$, $p_2 = 10^{-3}$, $p_3 = 10^{-3}$, and $30$ qubits.
    }
    \label{fig:surfaceQEC}
\end{figure}

Finally, we highlight the improved scaling with $k$ by analyzing $k$-uniform state preparation using the surface code circuits from Figure~\ref{fig:rot_surface_eg} and incorporating quantum error correction. Figure~\ref{fig:surfaceQEC} compares the hybrid scheme's performance for $d=3$ and $d=5$ surface codes. While the hybrid scheme maintains consistent performance with increasing $k$, purely physical preparation degrades with $k$.

\section{Discussion}\label{sec:discussion}

Our work demonstrates a scalable, fault-tolerant method for preparing encoded $k$-uniform states. We introduce a stabilizer tableau-based technique to determine $k$-uniformity and design fault-tolerant Clifford circuits for state preparation. Leveraging these states, we explore a hybrid physical–logical approach that balances some error protection with gate efficiency, showing its advantages over purely physical implementations for resource-state preparation. By extending $k$-uniform states into the logical space, our work paves the way for their use in key applications on physical quantum hardware.

Several promising directions emerge from this work. One avenue is to develop end-to-end protocols that utilize logical $k$-uniform states for specific applications, such as the various cryptographic applications~\cite{helwig2012absolute, helwig2013absolutely, shi2021k}. Validating these protocols in real-world settings would not only confirm their theoretical benefits but also demonstrate their practical value. Another research direction involves exploring logical $\Delta$-approximate $k$-uniform states, which can be prepared using shallower circuits than exact $k$-uniform states. Comparing their performance against exact states could offer valuable insights into their feasibility and usefulness on noisy quantum devices.

Finally, our hybrid scheme highlights additional opportunities for investigation. Here, we focus on a single unencoding step, but an important question is whether iterating multiple rounds of encoding and unencoding can continue to provide an advantage. Identifying the scenarios in which such a repetitive approach is most beneficial, and whether it holds practical relevance, remains an open challenge. In parallel, it is natural to compare this scheme against fully fault-tolerant methods or other $\mathcal{O}(p)$ error techniques for implementing arbitrary angle rotations, in order to pinpoint which regimes each method best serves.

\begin{acknowledgments}

We would like to thank Nishad Maskara and Madelyn Cain for many useful discussions. This work received support in part by the Banting Postdoctoral Fellowship and NSF QLCI grant OMA-2120757. DH is grateful for support from the Simons institute for the Theory of Computing, supported by DOE QSA. 

\end{acknowledgments}

\bibliography{refs}
\bibliographystyle{apsrev4-2}

\begin{appendices}

\renewcommand{\thesection}{\Alph{section}}
\renewcommand{\thesubsection}{\Alph{section} \arabic{subsection}}
\renewcommand{\thesubsubsection}{\Alph{section} \arabic{subsection} \roman{subsubsection}}
\def\theequation{\thesection\arabic{equation}}

\section{Method summary} \label{app:methodology}

We present an overview of our full method for identifying encoded $k$-uniform states. Before doing so, we describe the subroutine used repeatedly to identify the type of $k$-uniformity of a state. 

\begin{enumerate}[noitemsep]
\item Construct the binary symplectic representation of the stabilizer states’s generators.
\item Compute the rank (mod 2) for every combination of $\binom{N}{k}$ columns of the binary symplectic matrix, or of every combination whose column indices differ by no more than $\alpha$ if finding $\alpha$-separated $k$-uniformity. Record all resulting rank values.
\item Let $\tilde{r}$ be the minimum of all recorded rank values, and define $r = 2k - \tilde{r}$. Return $\Delta = 2 - 2^{1-r}$.
\end{enumerate}
We adopt the notation from mathematics or computer science that ``$\leftarrow$'' indicates a variable is being updated with a new value. Our method is:
\begin{enumerate}[itemsep=0pt]
    \item \textbf{Specify initial parameters:}
    \begin{enumerate}[itemsep=0pt]
        \item \label{step:specifyparams} 
        Select a quantum error-correcting code and list its transversal gates. The code's $\kappa$ and gates determine the circuit architecture (see Fig.~\ref{fig:kuniprep})
        \item Define the target $k$, $\Delta$, and the initial state.
        \item Set the maximum qubit count $N_{\rm max}$ for the search. Start with $N$ equal to the smallest integer strictly greater than $2k$ that fits the chosen architecture.
    \end{enumerate}
    \item \textbf{Search for $\Delta$-approximate $k$-uniform circuits:} \label{step:beginsearch} Determine the minimum circuit depth $\beta$ from a light cone argument.
    \begin{enumerate}[itemsep=0pt]
        \item Set the current sample index $\zeta = 0$. \label{step:zeta_int}
        \item  Identify a set of translationally invariant time-steps.  \label{step:symmetricstep}
        \item Enumerate all circuits of depth $\beta$ consistent with Step~\ref{step:specifyparams}, labeling them $0$ to $Z$. Circuits with lower indices prioritize translationally invariant layers, with boundary modifications introduced progressively. 
        \item \label{step:calc_kuni}  For each circuit $\zeta$, calculate $\alpha$-separated $\Delta$-approximate $k$-uniformity. If $\Delta$ meets the target, include the circuit and state in $\tilde{\Theta}_n$. 
        \item Increment $\zeta$ and repeat until all circuits are tested. 
        \item Refine $\tilde{\Theta}_n$ by calculate the $\Delta$-approximate $k$-uniformity. If $\Delta$ equals the target $\Delta$, include the circuit and state in the set $\Theta_n$. 
        \item If no circuits are found, increment $\beta \leftarrow \beta + 1$ and return to Step~\ref{step:beginsearch}
    \end{enumerate}
    \item \textbf{Repeat for larger $N$:} If $N < N_{\rm max}$, increment $N \leftarrow N + 1$ and return to Step~\ref{step:zeta_int}.
\end{enumerate}

At the end of this procedure, one has a set of circuits $\Theta_N$ for different values of $N$. To find a repeatable pattern, one can then identify a circuit architecture that remains consistent across all $N$. Next, verify that this architecture continues to prepare $k$-uniform states as $N$ increases beyond $N_{\rm max}$, ensuring that the structure holds for all $N$ with high confidence. Using this procedure, we obtain the circuits described in Sec.~\ref{sec:kuniprotocol}.

\section{Example of calculating the \texorpdfstring{$k$}{k}-uniformity}\label{app:kunicalc_example}

We illustrate our approach for calculating the $k$-uniformity with a simple example. Here we use the $\ket{\overline{0}}$ stabilizer state of the five-qubit code. The stabilizer generators $G(\mathcal{S})$ for $\ket{\overline{0}}$ can be represented using the binary symplectic representation as follows: 
\begin{align}
 \left[\begin{array}{ccccc|ccccc}
 1 & 0 & 0 & 1 & 0 & 0 & 1 & 1 & 0 & 0\\
 0 & 1 & 0 & 0 & 1 & 0 & 0 & 1 & 1 & 0\\
 1 & 0 & 1 & 0 & 0 & 0 & 0 & 0 & 1 & 1 \\
 0 & 1 & 0 & 1 & 0 & 1 & 0 & 0 & 0 & 1 \\
 0 & 0 & 0 & 0 & 0 & 1 & 1 & 1 & 1 & 1
 \end{array}\right]
\end{align}
To trace out certain qubits, we remove the columns corresponding to the qubits in $\bar{A}$. 
For $k = 2$, we can choose $\bar{A} = \{1, 3, 5\}$, leaving: 
\begin{align}
 \left[\begin{array}{cc|cc}
 0 & 1 & 1 & 0\\
 1 & 0 & 0 & 1\\
 0 & 0 & 0 & 1\\
 1 & 1 & 0 & 0\\
 0 & 0 & 1 & 1
 \end{array}\right].
\end{align}
This matrix represents the stabilizers restricted to region $A$. To determine how many stabilizers are independent, and thus compute $I_A$, we calculate the rank of this matrix mod 2. In this case, $I_A = 4$. We then repeat this calculation for all other regions of size $k$. Since $\Delta$ decreases as $I_A$ increases, we identify the minimum $I_A$. Using this value, we compute $\Delta$. For all subsets $A$ with $\abs{A} = 2$, the calculation gives $I_A = 4$. Therefore, the five-qubit code is $k = 2$-uniform, with an approximation $\Delta = 0$.

\section{Decay of \texorpdfstring{$\Delta$}{Delta}-approximate \texorpdfstring{$k$}{k}-uniform states} \label{app:deltaapprox}

\begin{figure}[h]
 \centering
 \includegraphics[width=\columnwidth]{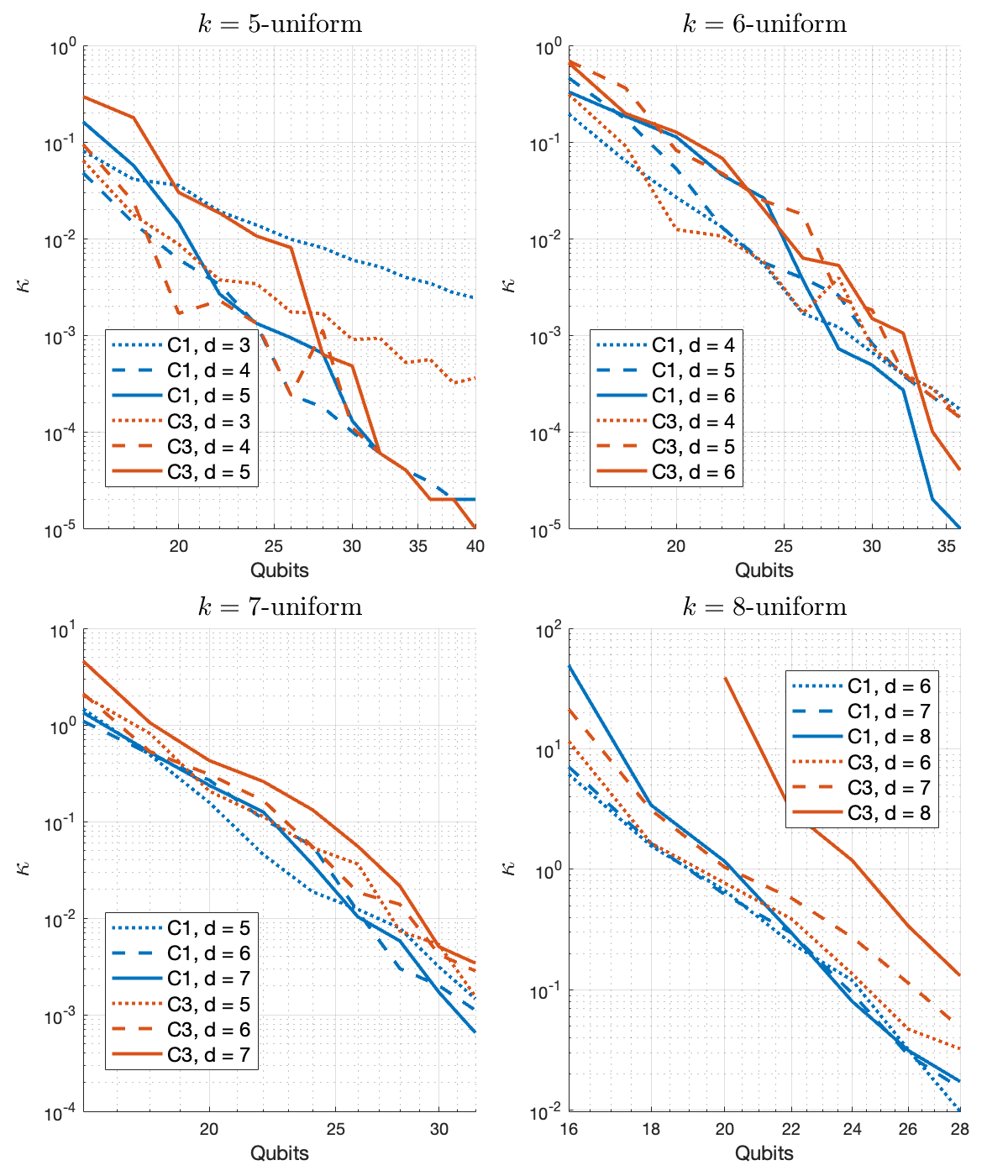}
 \caption{
 \caphead{Decay of non-$k$-uniform states.} $\kappa$ denotes the ratio of $\Delta$-approximate $k$-uniform subsets to those that are exactly $k$-uniform. Circuits $C_1$ and $C_3$ correspond to layers identical to one time step of the circuits shown in Fig.~\ref{fig:k1} and Fig.~\ref{fig:k3}, respectively. The circuit depth is denoted by $d$. The relative number of non-$k$-uniform states exhibits an approximate power-law decay with increasing circuit size.
 }
 \label{fig:decay}
\end{figure}

We can characterize $\Delta$-approximate $k$-uniform states by their maximal deviation $\Delta$ and distribution across all $k$-site combinations. As $n$ grows, the fraction of $\Delta$-approximate $k$-uniform subsets relative to exact $k$-uniform ones decreases in an approximate power-law (See Fig.~\ref{fig:decay} of Appendix). Notably, we see that simply adding more circuit layers does not prevent this decline.

\end{appendices}

\end{document}